\documentclass[manuscript,,screen,anonymous = FALSE,,review = FALSE]{acmart}
\usepackage[normalsize,FIGTOPCAP]{subfigure}
\usepackage{float}
\usepackage{enumitem}

\AtBeginDocument{%
  \providecommand\BibTeX{{%
    \normalfont B\kern-0.5em{\scshape i\kern-0.25em b}\kern-0.8em\TeX}}}

\copyrightyear{2022}
\acmYear{2022}
\setcopyright{acmcopyright}
\acmConference[LAK22]{LAK22: 12th International
Learning Analytics and Knowledge Conference}{April 12--16, 2021}{Irvine, CA, USA}
\acmBooktitle{LAK22: 12th International Learning Analytics and Knowledge
Conference (LAK22), April 12--16, 2021, Irvine, CA, USA}
\acmPrice{15.00}
\acmDOI{10.1145/3448139.3448150}
\acmISBN{978-1-4503-8935-8/21/04}

\begin{document}
\title[Analysis of Behavioral Patterns in an Online Learning Environment]{A Multi-level Trace Clustering Analysis Scheme for Measuring Students' Self-Regulated Learning Behavior in a Mastery-Based Online Learning Environment}

\author{Tom Zhang}
\email{Tom.Zhang@ucf.edu}
\affiliation{%
  \institution{University of Central Florida}
  \streetaddress{4111 Libra Dr.}
  \city{Orlando}
  \state{Florida}
  \country{USA}
  \postcode{32816}
}

\author{Michelle Taub}
\email{Michelle.Taub@ucf.edu}
\affiliation{%
  \institution{University of Central Florida}
  \streetaddress{4111 Libra Dr.}
  \city{Orlando}
  \state{Florida}
  \country{USA}
  \postcode{32816}
}

\author{Zhongzhou Chen}
\email{Zhongzhou.Chen@ucf.edu}
\affiliation{%
  \institution{University of Central Florida}
  \streetaddress{4111 Libra Dr.}
  \city{Orlando}
  \state{Florida}
  \country{USA}
  \postcode{32816}
}

\renewcommand{\shortauthors}{Zhang, et al.}

\begin{abstract}
The study introduces a new analysis scheme to analyze trace data and visualize students' self-regulated learning strategies in a mastery-based online learning modules platform. The pedagogical design of the platform resulted in fewer event types and less variability in student trace data. The current analysis scheme overcomes those challenges by conducting three levels of clustering analysis. On the event level, mixture-model fitting is employed to distinguish between abnormally short and normal assessment attempts and study events. On the module level, trace level clustering is performed with three different methods for generating distance metrics between traces, with the best performing output used in the next step. On the sequence level, trace level clustering is performed on top of module-level clusters to reveal students' change of learning strategy over time. We demonstrated that distance metrics generated based on learning theory produced better clustering results than pure data-driven or hybrid methods. The analysis showed that most students started the semester with productive learning strategies, but a significant fraction shifted to a multitude of less productive strategies in response to increasing content difficulty and stress. The observations could prompt instructors to rethink conventional course structure and implement interventions to improve self-regulation at optimal times.
\end{abstract}



\maketitle

\section{Introduction}

Studying students' self-regulated learning (SRL) behavior  in online learning environments through analysis of trace data has become one of the focuses of learning analytics research, especially since the COVID pandemic forced educators around the world to abruptly switch to online learning. SRL identifies students as playing active roles in their learning by engaging in planning, monitoring, and reflection strategies during learning \cite{Winne2018, Zimmerman2013}, which is imperative for online learning as students are less reliant on instructors for in-the-moment assistance and are required to engage in more independent, structured, and regimented learning to keep up with class activities and content mastery. A number of recent studies have employed multiple data mining techniques such as sequence pattern analysis, process mining, and hierarchical clustering to identify, visualize and compare students' SRL strategies from trace data \cite{Maldonado-Mahauad2018, Fan2021}, with the latest example being the Trace-SRL framework developed by Saint et. al. \cite{Saint2020}. The Trace-SRL framework first maps event sequences onto micro-SRL processes based on an SRL model \cite{Zimmerman2011}, then identifies SRL strategies by conducting two levels of sequence clustering on SRL action traces, and finally creates process models of SRL learning strategies for different groups of students.  

The data set analyzed in Saint et. al. contains a rich variety of events from attempting problems to accessing the scheduling page or the dashboard. The online learning environment also provided students with a relatively high level of freedom to access different course contents in their preferred order. Under those conditions, different micro-level SRL actions will likely result in distinct event sequences, which allowed researchers to create a "dictionary" that uniquely maps event sequences onto SRL actions. The relatively high variability in event sequences also enables the dissimilarity between different traces to be calculated based on transition rates and state frequencies observed in the data set, which is the default option provided in mainstream trace clustering tools \cite{TraMineR}.

However, online learning systems or online learning data sets are oftentimes more restrictive, meaning that they provide fewer event types and less variability in event sequences. This could happen for several reasons, first, certain restrictions could be beneficial for learning from a pedagogical standpoint. For example, in a mastery-based learning system designed based on the principles of deliberate practice \cite{Ericsson1993}, students would be required to complete a sequence of tasks in a pre-determined order. From a "preparation for future learning" perspective \cite{Schwartz2005}, requiring students to attempt a problem first could improve the level of learning from subsequent study of the instructional materials. Second, online homework or intelligent tutoring systems are often integrated into an online course hosted on a separate learning management system such as Canvas. In those cases, data from the sub-system may only contain a subset of events such as attempts and page access, whereas other types of events such as checking the dashboard are stored in a different data set that may not be available to the researcher.
		
For those more restrictive date sets, direct application of existing analysis schemes faces two prominent challenges. First, it will be challenging to reliably identify one-to-one correspondence between event sequences and microscopic SRL strategies. This is because on one hand, students adopting different learning strategies may end up producing similar or even identical event sequences due to restrictions imposed by the system. On the other hand, each event or event sequence may be reflective of more than one SRL actions. For example, if a student accessed the learning material after a failed problem-solving attempt, the decision could indicate that the student reflected on the previous attempt, but may also be part of the students' initial plan to learn the skills needed to solve the problem, or a combination of both. Second, sequence clustering based entirely on the frequencies of events in the data set could be insufficient in identifying key differences in students' learning strategies in a restricted online system. For example, a student who first attempted to solve a problem and then guessed on subsequent attempts likely had different strategies than those who started guessing on attempt one. Yet there is no guarantee that the frequency differences contained in the data itself will be sufficient to separate those two states, especially since the variability in trace sequences in a restricted system is markedly smaller. Both of those challenges are present in the analysis of all types of online learning trace data sets but are more prominent in limited data sets from restrictive systems.

Therefore, in this study we will try to answer the following questions: can we adopt and modify similar strategies and algorithms such as those used in the Trace-SRL framework to analyze students' learning data from a restrictive learning system? To what extent can this approach still provide valuable insight into students' self-regulation strategies, learning behaviors, as well as goals and motivations, without the need for directly mapping event sequences onto unique SRL actions? 

Below, we will first briefly introduce the design of the mastery-based online learning modules (OLM) system being analyzed in the current study. We then outline our attempt to develop an analysis scheme suitable for trace data produced from the OLM system, by employing three different clustering algorithms on three consecutive levels of data granularity.

\subsection{Design of OLM and OLM Sequences}
\begin{figure}[!h]
    \includegraphics[width=.45\textwidth]{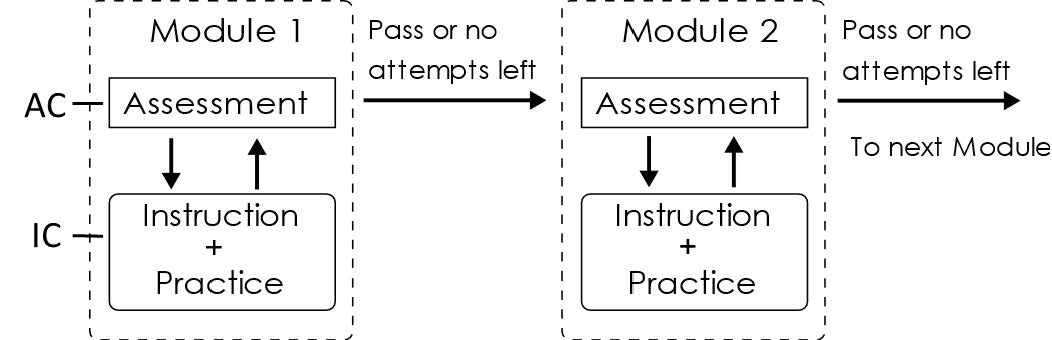}
    \caption{Schematic illustration of OLM design, adapted from \cite{Chen2020}}
    \label{fig: OLM_design}
    \Description{Design diagram of the Online Mastery-Based Learning Modules.}
\end{figure}

Designed based on principles of mastery-learning \cite{Gutmann, Bloom1968} and deliberate practice \cite{Schwartz2005}, an OLM combines assessment, instruction, and practice into a single online learning unit. Each OLM is focused on explaining one or two basic concepts, or developing the skills to solve one kind of problem, designed to be completed in 5 to 30 minutes. Each individual OLM consists of an assessment component (AC) which tests students' content mastery in 1-2 questions, and an instructional component (IC) with instructional text and practice problems on the topic (see Figure \ref{fig: OLM_design}). Upon accessing a module, students are shown the learning objectives of the current module and are required to make an initial attempt on the AC before being allowed to access the IC. Students can make additional attempts on the AC at any time after the first attempt and are not required to access the IC. This design is motivated by both the "mastery-learning" format that allows students who are already familiar with the content to proceed quickly to the next assignment and by the concept of "preparation for future learning" intending to improve students' learning from the IC by exposing them to the questions first. It also allows researchers to more easily measure knowledge transfer between subsequent modules \cite{Whitcomb2021}.

A number of OLM modules form an OLM sequence on a more general topic (e.g., conservation of mechanical energy) and students are required to pass the AC or use up all attempts before moving onto the next OLM in that sequence. A typical OLM sequence consists of 5-12 modules and are assigned as self-study homework for students to complete over a period of one to two weeks. The reader can preview all the OLMs at https://canvas.instructure.com/courses/1726856.

\subsection{Analysis of OLM Trace Data via Multi-Level Clustering}
The current analysis scheme involves three consecutive clustering operations performed on the trace data from the OLM system at the event level, module level, and sequence level.

\subsubsection{Level I: Clustering of Individual Events}
Prior research on OLMs has shown that students' use of learning strategies may not be reflected by the order of events, but rather by the quality of events \cite{Chen2020, Garrido2020}. More specifically, an abnormally short assessment attempt is likely the result of random guessing or answer copying and may be indicative of the student adopting a performance-avoidance goal. Therefore, the main goal of the event level clustering is to distinguish between abnormally short guessing attempts and longer problem-solving attempts. This is achieved by fitting the log-duration distribution with finite mixture modeling (FMM) \cite{mixsmsn}, which can be seen as clustering based on a single continuous variable (event duration). The same technique is also applied to identify and exclude very short study events which likely originated from a student clicking through the instructional contents without meaningfully interacting with them.

\subsubsection{Level II: Clustering Interaction Traces on Individual Modules} 
To identify the main strategies that students adopt when interacting with individual OLM modules, we partition students' event traces on a single module into multiple "module-level clusters", or m-clusters, using hierarchical agglomerative clustering. We will test and compare three different techniques for determining the dissimilarity between different traces: 

\begin{enumerate}
    \item A purely data-based strategy using optimal matching distance similar to the method used in \cite{Saint2020, Jovanovic2017}, and relying solely on frequencies observe from the data.
    \item A hybrid strategy with which part of the dissimilarity calculation is informed by the design of the platform.
    \item A purely theory-based strategy that determines the dissimilarity based on a set of features that are deemed to be important in the context of an OLM, informed by an SRL framework adapted for restrictive learning systems.
\end{enumerate}

For each resulting m-cluster, a process map is generated via exploratory process mining using the package bupaR \cite{bupaR} with the top 60\% of most frequent traces, as an intuitive visual representation of the main characteristics of each interaction strategy.

\subsubsection{Level III: Clustering of Interaction Traces on OLM Sequences}
As a result of module-level clustering, students' use of learning strategies for each OLM sequence can be captured as a trace of m-cluster memberships. We can then partition those traces into sequence-level clusters, or s-clusters, to capture the main strategies employed by each student to interact with the entire OLM sequence. In Trace-SRL, student-level clustering is based on the frequency of adopting different strategies using the Euclidean distance. In the current analysis, the s-clusters are identified based on the optimal matching distance between traces calculated using the TraMineR package \cite{TraMineR}. Using the optimal matching distance preserves the temporal information in trace data, which will enable us to investigate when students change their interaction strategies in response to different factors such as content difficulty.

Due to the computational complexity of this analysis scheme, the current study will analyze trace data collected from three representative OLM sequences selected from the beginning, middle and end of the semester, to answer the following three research questions:

\begin{enumerate}[label=RQ\arabic*:, left = 10pt]
    \item Which of the three clustering strategies (Data, Hybrid, Theory) will produce the most well organized and interpretable m-clusters?
    \item What are the main strategies that students adopted to interact with OLMs, and how are those strategies related to their assessment performance?
    \item How do students' choice of strategies change over time, and how can we interpret their use of strategy in the framework of SRL?
\end{enumerate}

\section{Methods}\label{sec: Methods}
\subsection{OLM Design and Implementation}
The OLM modules are created and hosted on the Obojobo learning objects platform, an open-source online learning platform developed by the Center for Distributed Learning at the University of Central Florida \cite{CenterforDistributedLearning}. In the current iteration, the assessment component of each OLM contains 1-2 multiple choice problems and permits a maximum of 5 attempts. The first 3 attempts are sets of isomorphic problems assessing the same content knowledge with different surface features or numbers. On the 4\textsuperscript{th} and 5\textsuperscript{th} attempts, students are presented with the same problems in the 1\textsuperscript{st} and 2\textsuperscript{nd} attempts respectively and are awarded 90\% of credit. The instructional component of each module contains a variety of learning resources including text, figures, videos, and practice problems. Each OLM sequence contains 3-12 OLMs, which students must complete in the order given, with completion defined as either passing the assessment or using up all 5 attempts. Readers can access example OLMs at https://canvas.instructure.com/courses/1726856.

\subsection{Instructional Conditions and Data Collection}
Data used in this study was collected from a calculus-based university introductory physics course taught in the Fall 2020 semester. A total of 251 students were initially enrolled in the class, of which 67 were females, and 98 were historically underrepresented minorities. The course was taught asynchronously using pre-recorded lecture videos as the main method for content delivery as a result of the COVID pandemic. Students and instructors interacted via messages, posts, and short video conferences via Microsoft Teams (for more information on the course design, see https://www.aaas-iuse.org/resource/course-design/). Students were required to take a total of 7 quizzes during the semester, with each quiz lasting 20 minutes and proctored by a teaching assistant. 

A total of 70 OLMs consisting of 9 sequences were assigned as the only form of online homework. Each OLM sequence was assigned for students to complete over 1-2 weeks, with all OLMs in a sequence due on the same day. The OLM modules accounted for 24\% of course credit, with no additional homework assignments. Students could earn extra credits by completing some OLMs earlier than the due date, as explained in more detail in \cite{Felker2020}.

\subsection{Collection and Analysis of Student Log Data}
Students' clickstream log data collected from the Obojobo platform is first processed into attempt events and study events. An attempt event starts when the student enters the assessment page of the module, and ends when the student clicks the submit bottom on the assessment page. During this period the student is unable to navigate to any other pages in the current module or to other modules. The duration of the attempt event is the time between those two clicks minus the duration of 1) when the browser window is either closed or minimized, or when another window is in focus and 2) any non-active duration beyond 10 minutes. An attempt event is labeled as "pass" only if the student correctly answers all questions in the assessment, and fail otherwise. 

A study event starts when the student clicks on any page in the instructional component of the module, and ends when the student clicks on the last record before a new attempt event is initiated, or the last record for the student on the module. In other words, a study event includes all the interaction with the instructional component between two attempt events. The duration of the study event is calculated as the sum of all the time spent interacting with each instructional page, minus the duration of inactive periods as explained above. 

For the current study, we selected event data from Sequence 1: Motion in 1 Dimension, Sequence 6: Mechanical Energy, and Sequence 9: Angular Momentum. The three sequences consist of a total of 26 modules, and the resulting data set contains a total of 5960 traces. In addition, all records after the first passing attempt or after the last attempt were truncated for simplicity of analysis, since there were significantly fewer records after passing or using up all attempts, and most of those events took place before an exam \cite{Chen2020}. The topics of the three sequences are sufficiently different so that learning outcome of earlier sequences should not significantly impact the learning behavior of later sequences. 

\subsection{Clustering Level I: Categorization of Events}
At level I, abnormally short attempts on the AC of a give OLM are distinguished from normal AC attempts, by fitting the log duration distribution of all attempts on a single module using FMM. FMM is a model-based clustering algorithm that divides a population into subgroups according to one or more observable characteristics, by fitting the distribution of characteristics with a finite mixture of normal or skewed probability distributions. When two or more distinct problem-solving behaviors are present, the log attempt duration distribution can be fitted with the sum of two or more distributions, with the shortest distribution corresponding to abnormally short attempts.
In the current study, we fit the log-duration of each assessment attempt using either normal or skewed distribution models using the R package mixsmsn \cite{mixsmsn}, following the fitting procedure described in detail in a previous study \cite{Chen2020}. 
In the case when a single component distributed is the best fit for the duration, then the cutoff is set as either 2 standard deviations below the mean duration, or 15 seconds, whichever is longer. This is because a previous clinical study indicated that attempts under 15s are likely to arise from complete random guessing \cite{Guthrie2020}. 

We also conducted mixture-model fitting of the combined log duration of all study events from all modules in the data set, to determine the cutoff time between normal study events and very short study events that are likely the result of students clicking through the instructional pages. Study events that are shorter than the cutoff are excluded from the data set since those events are more likely the result of accidental or unintentional clicks rather than a deliberate decision to engage with the material.

\subsection{Level II: Module-Level Trace Clustering}
As a result of Level I clustering, each students' interaction with a given OLM is represented by a trace of either normal or short attempt events and study events that are longer than the minimal duration. Each attempt is treated as a separate event and labeled as "Attempt\_N" with N being the attempt number. Short attempts are labeled as "Attempt\_N\_S" to distinguish them from normal attempts. For example, a trace of \{Attempt\_1\_S, Study, Attempt\_2, Attempt\_3\} indicates that the student took 3 attempts on the OLM, with the first attempt being a short attempt, and took a study session (longer than the minimum cutoff) between attempts 1 and 2. 

Hierarchical agglomerative clustering using Ward's method is performed via the R package cluster \cite{cluster} on traces from all three selected OLM sequences, with each trace treated as a data entry. We investigate and compare three different kinds of strategies for calculating the distance between two traces: purely data-driven strategies, hybrid strategies, and purely theory-based strategies, explain in detail below. The resulting module-level clusters, or m-clusters, reflect different learning strategies that students adopt to interact with individual OLMs.

\subsubsection{Data-Driven Trace Distances}
The distance metric used for data-driven strategy is the Levenshtein distance, defined as the minimum number of operations needed to transform a given string $a$ into another $b$ via single letter operations (insertion, deletion, and substitution). The matrices generated by the cost generation methods determine the relative weight of substitution between letters in the alphabet\footnote{Here, alphabet refers to the available range of letters in the state space.}, while the indel costs refer to the relative weight of insertion/deletion of letters in the alphabet. 

There exists three cost generation methods native to the TraMineR package; the INDEL, TRATE, and FUTURE methods which utilize \textit{only} information observed from the data. The INDEL method generates the cost matrix by calculating the relative frequency of each state, the TRATE method by transition rates between adjacent states in a given trace, and the FUTURE method by the Chi-Square distance for the state probability vectors. A more detailed description of the methods can be found in the documentation of the TraMineR package \cite{TraMineR}. All three methods are being tested for the current analysis, and the one that generates the most well-formed clusters (explained in more detail below) is selected as the result of the pure data-driven strategy.

\subsubsection{Hybrid Trace Distance}
The local structure in the TRATE most closely matches the theorized actions associated with SRL behaviors out of the three data-based methods. As such it requires the least modification of the dissimilarity inputs to better match and subsequently separate such predicted behaviors. To do this, we constructed a vector of the insertion/deletion costs of the alphabet with the following constraints in mind:

\begin{enumerate}
    \item Study events after Attempts 1, 2, and 3 are significantly different from study events after Attempts 4 and 5.
    \item Brief Attempts are significantly different from Normal Attempts.
    \item The first Brief Attempt is significantly different from the first Normal Attempt.
\end{enumerate}

\subsubsection{Theory-Based Trace Distance}

For each OLM, students are presented with two tasks: a required task to solve the problem in the assessment component, and an optional task to study the learning material. Therefore, students will need to make two types of decisions:

\begin{enumerate}
    \item Whether to seriously engage in problem solving on a given attempt (resulting in a normal length attempt) or to make a guess (usually resulting in a short attempt).
    \item Whether to engage with the study material if the first attempt fails.
\end{enumerate}

Using the three macro-level self-regulatory phases presented in Zimmerman’s model of SRL \cite{Zimmerman2000}, we propose six features that capture students’ interactions with the OLMs and their associated SRL processes, based partly on two earlier studies \cite{Chen2020, Zhang2021}:

\begin{enumerate}
    \item Total Number of Assessment Attempts (nA): This reflects the quality of \textbf{performance} on both the problem-solving task and the studying task. In general, passing on fewer attempts reflects either high incoming knowledge or successful learning or both.
    \item Number of Attempts before Study\footnote{If a student never accessed the instructional materials, this feature has a value of 5, which is equivalent to having studied after the 5\textsuperscript{th} attempt. Since event logs after the passing attempt of 5\textsuperscript{th} attempt are excluded from the current data set, a value of 5 always represents a no-study trace. The choice of coding amplifies the decision of whether to access the study materials when interacting with the OLMs.} (nY): When or whether to access learning materials can be influenced by either \textbf{planning} or \textbf{reflection}. While accessing the learning materials after the first mandatory attempt is more likely the result of planning, accessing after multiple failed attempts is more likely the result of reflecting on previous attempt performance. 
    \item Fraction of Short Attempts Among All Attempts (fS): Since most short attempts likely originate from either guessing or copying behaviors, a high fraction of short attempts could indicate low prior knowledge, low self-efficacy, low effort, or limited execution strategies like time management.
    \item Is the First Attempt Short (1S): The first attempt is of particular significance as it reflects students' \textbf{planning} of the time and effort to spend on the mandatory first attempt. A short first attempt can be a sign that the student is trying to conserve time and effort on the task by guessing. It could also be influenced by the student's \textbf{reflection} on experience in the previous modules and their perceived self-efficacy within the course. A student with lower confidence, or is aware of their low prior knowledge, will be more likely to make a short first attempt to access the learning materials.
    \item Is the Last Attempt Short (lS): The last attempt is also of particular significance since it is the passing attempt in all event traces, except for those with five failed attempts. A short final passing attempt signifies lower \textbf{performance} on both learning and problem-solving tasks, and probably lower levels of \textbf{reflection} on learning.
    \item Did the Student Abort the Module (Ab): This feature represents a small number of (22 out of 5960) event traces that ended on a failed attempt that is not attempt 5 from the rest of the event traces. Those traces exist either because the student aborted the module, or because of corrupted data logs.
\end{enumerate}

It is worth noting that the number of times a student accessed the instructional materials is not included as a feature. This is because, in 90\% of cases, students interact with all the learning materials at once and in most other cases, one study event is significantly longer than the others. It is also much less clear whether a higher degree of access reflects higher or lower quality of learning.

 
Since features 1, 2, and 3 are numeric while features 4, 5, and 6 are binary, the distance metric between two event traces is computed using the Gower dissimilarity coefficient. We tested multiple sets of feature weights for calculating the Gower coefficient. The best cluster structure, as explained below, was produced when the weights for nA and lS are set to 0.5 and all other weights set to 1.0, which emphasizes the forethought phase.

\subsubsection{Determining the optimal number of m-clusters}
Since agglomerative clustering produces a tree structure of all possible numbers of clusters, we choose to determine the optimum number of m-clusters based on the average silhouette value of each cluster. A silhouette is a measure of the ratio of intra- and inter-cluster variability which is described in \cite{Silhouette}. The term "average silhouette" refers to the average silhouette value for every point in the clustering algorithm. Theoretically, the optimal number of clusters is chosen to maximize the average silhouette, as it indicates that the variability within clusters is very small compared to the variability between clusters, thus being well defined. 

However, in practice, the current data set of 5960 traces contains only 53 unique traces. As a result, the average silhouette will always reach the global maximum at or near 53 clusters. Several clustering strategies resulted in a local maxima of average silhouette in under 10 clusters, indicating a relatively well-defined cluster structure with a small number of clusters. In those cases, the number of m-clusters is chosen using the local maximum average silhouette.

\subsubsection{Visualizing m-Cluster Structure using Process Maps}
The main characteristic of each m-cluster is visualized by creating process maps (PMs) using the R package processmapR \cite{processmapR} for event traces in each cluster. A "PASS" event is appended to the end of the trace if the last attempt is a passing attempt. Otherwise, the sequence ends with the last attempt. The top 60\% of most frequent event traces in each m-cluster are included in the map to reduce "spaghetti effects" caused by rare traces. 

\subsection{Level III: Sequence-Level Trace Clustering}
Following module-level clustering, a student's event trace interacting with each OLM is classified as belonging to one of several m-clusters. Thus, the student's interaction with an OLM sequence of n modules can be captured by a sequence-level trace of n elements in the form of $\{m_1,m_2,...,m_n\}$, with each element $m_i$  being a number representing the m-cluster that the student’s event trace on module $i$ belongs to. 

We again perform hierarchical agglomerative trace clustering on the sequence-level traces for each of the three OLM sequences separately using the Ward method. The distance between two traces is calculated using the optimal matching distance via the TRATE method, as it takes into account the local ordering of states. Since each student contributes one trace per sequence to the data set, the s-clusters are a reflection of the strategy adopted by individual students.

The number of s-clusters for each sequence is determined by the maximal average silhouette value between 2 and 10 clusters. In the case that the maximum average silhouette is 2 clusters, but a local maximum exists for a higher number of clusters, then the higher number of clusters is selected in order to better elicit relatively rare but distinct interaction strategies. 

\subsection{Correlation With Assessment Outcomes}

To explore the correlation between identified s-clusters and students' performance on assessment, we divided the students into three tertiles of near equal size along quiz score performance (High, Medium, and Low) following the practice in \cite{Jovanovic2017}. Quiz scores were chosen over exam scores since quizzes were being proctored and exams were not during the Fall 2020 semester. We then compared the frequency distribution of s-cluster membership between the three cohorts using Fisher's exact test. When a significant difference was observed for a given sequence, we then performed post hoc analysis via pairwise Fisher exact tests for every s-cluster between every two performance cohorts, using the Benjamini, Hochberg, and Yekutieli ("fdr") method \cite{fdr} for p-value adjustment.

\section{Results}\label{sec: Results}
\subsubsection{Event Level Clustering}
Of the 26 module assessments included in this data set, the attempt duration distribution of 8 assessments were fitted with 1 component FMMs, and the rest are all fitted with 2 or more components FMMs. For 4 assessments, the short versus normal cutoffs as determined by FMM modeling were less than 15s and were thus adjusted to 15s. Twenty-one modules had short versus normal cutoffs between 15 and 120s, and 2 modules had cutoffs beyond 120s. Assessments involving numerical calculation problems had longer cutoffs compared to those involving conceptual questions.

\subsection{Module-Level Trace Clustering}
\subsubsection{Comparison of Different Clustering Methods} The silhouette plots for each of the three methods (data-based, hybrid, and theory-driven) is shown below in figure \ref{fig: s-Silhouette}:

\begin{figure}[!h]
    \centering
    \includegraphics[width = .45\textwidth]{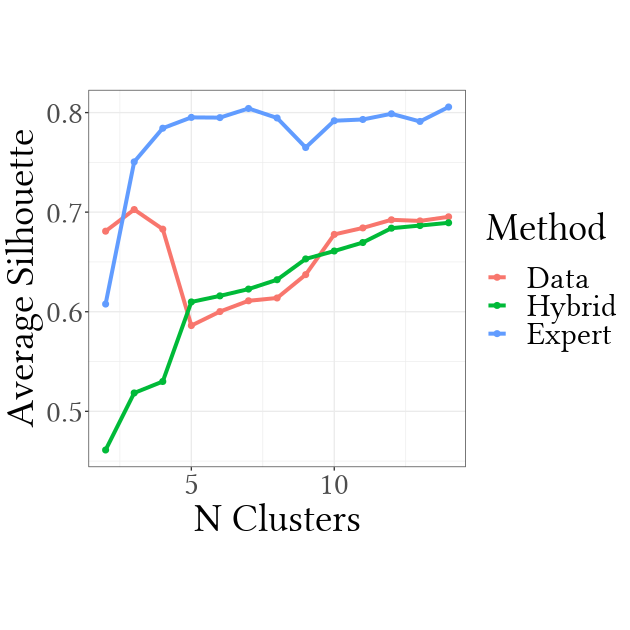}
    \caption{Plots of the Average Silhouettes for each distance generation method for m-Clusters}
    \label{fig: s-Silhouette}
\end{figure}

The theory-based method outperformed the other two methods with a local maximum average silhouette of over 0.8 at 7 clusters. The purely data-driven method had a local maximum of 0.7 at three clusters, while the average silhouette of the hybrid method did not reach a local maximum in under 10 clusters.

\begin{figure}
    \centering
    \subfigure[Pure Data Process Map]{
        \includegraphics[width = .75\textwidth]{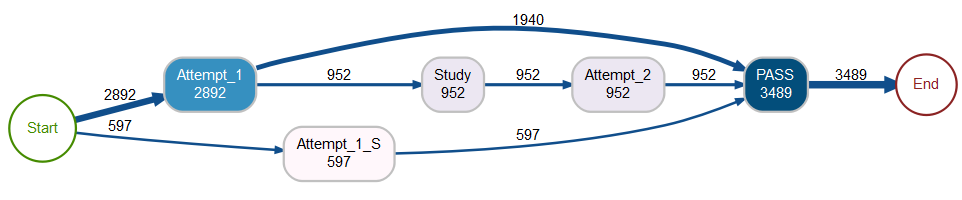}
    }
    \subfigure[Hybrid Process Map]{
        \includegraphics[width = .75\textwidth]{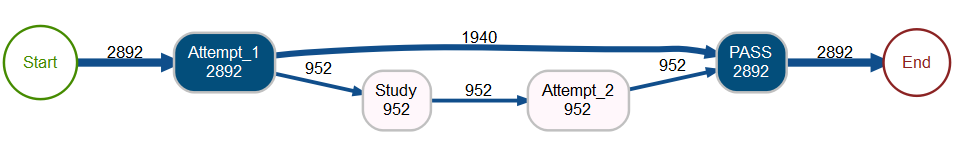}
    }
    \caption{Process maps generated from the (a) INDEL and (b) Hybrid cost matrix generation methods.}
    \label{fig: nonExpert PM}
\end{figure}

For all the clusters identified by either the pure data or the hybrid methods, the corresponding process maps (PMs) show that traces both with and without a study event are grouped into the same cluster, as shown in the example PM in Figure \ref{fig: nonExpert PM}. Furthermore, the pure data method also cannot distinguish between normal and short assessment attempts. Since both methods failed to distinguish features important for interpreting students' SRL strategies, we chose to only show one example PM for each method


\begin{figure}
    \centering
    \subfigure{
        \includegraphics[width = \textwidth]{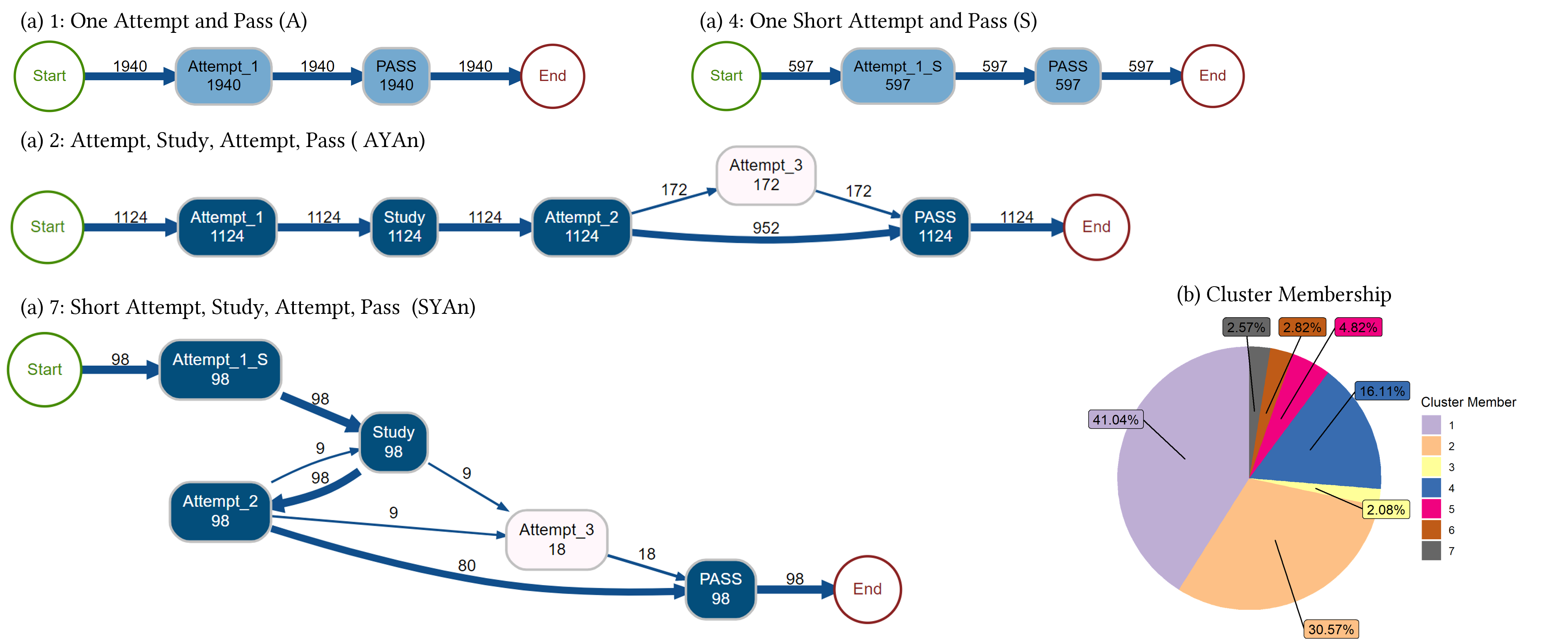}
    }
    \subfigure{
        \includegraphics[width = \textwidth]{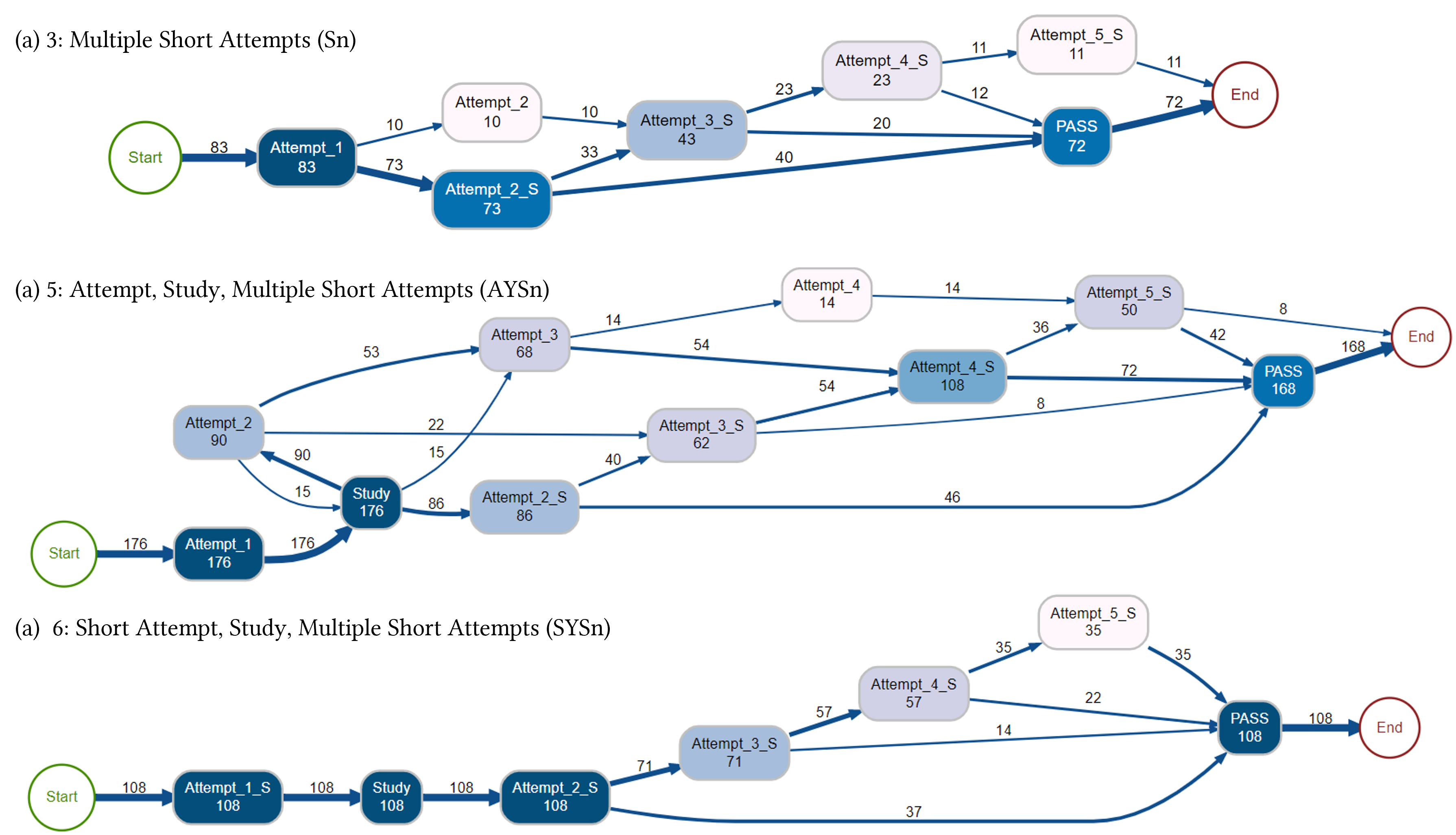}
    }
    \caption{(a) 1-7: The most common 60\% of traces in each m-cluster. The shorthand name for each PM captures its main characteristics, using the following convention: A, S, and Y represent an Attempt, a Short Attempt, and a Study event respectively. A letter followed by an n represents multiple occurrence of the event represented by the letter. (b): Pie chart showing the relative cluster membership of the student traces.}
    \label{fig: Expert PM}
\end{figure}

The PMs for the m-clusters generated by the theory-based method is shown in figure \ref{fig: Expert PM}, labeled using the shorthand notation explained in the figure caption. The PMs can be divided into two groups based on the "lengths" of the PMs: Traces in PMs 1, 2, 4, and 7 all reach the PASS state within 3 attempts. PMs 3, 5, and 6 all contain traces that pass on either the 4\textsuperscript{th} or 5\textsuperscript{th} attempt, or did not reach the pass state after the 5\textsuperscript{th} attempt.

In PMs 1, 2, 4, and 7, the majority of cases passed the module on the 1\textsuperscript{st} or 2\textsuperscript{nd} attempt, with a smaller fraction ($<20\%$) passing on the 3\textsuperscript{rd} attempt. As shown in the PMs, most traces in m-cluster 1 (A) passed the module without studying the learning material whereas those in m-cluster 2 (AYAn) passed the module in a few attempts after studying. m-cluster 7 (SYAn) is similar to m-cluster 2, except that the initial attempt is a short one. The passing attempts in all three PMs are all normal attempts. All of the traces in m-cluster 4 (S), on the other hand, passed the module on short attempts with no study event, and at least 60\% of the traces passed on only one short attempt, as shown in the PM. It is likely that the traces in m-cluster 4 (S) originated from the student either making a lucky guess or obtaining the answer from another source. 

A significant common feature of PMs 3, 5, and 6 is that all the passing attempts are short attempts. M-cluster 3 (Sn) started with a normal attempt, followed by multiple short attempts without a study event in most traces. In m-clusters 5 (AYSn) and 6 (SYSn), a study event followed either a normal or a short first attempt. In those three clusters, it is likely that students are guessing on the assessments after either having taken the first attempt (3) or after studying the instructional materials (5, 6).  

Since the theory-driven distance metric produced m-clusters with both better structure and better interpretability, we chose to use those results in Level III clustering analysis.

\subsection{Sequence-Level Clusters}

\begin{figure}[!ht]
    \centering
    \includegraphics[width = .95\textwidth]{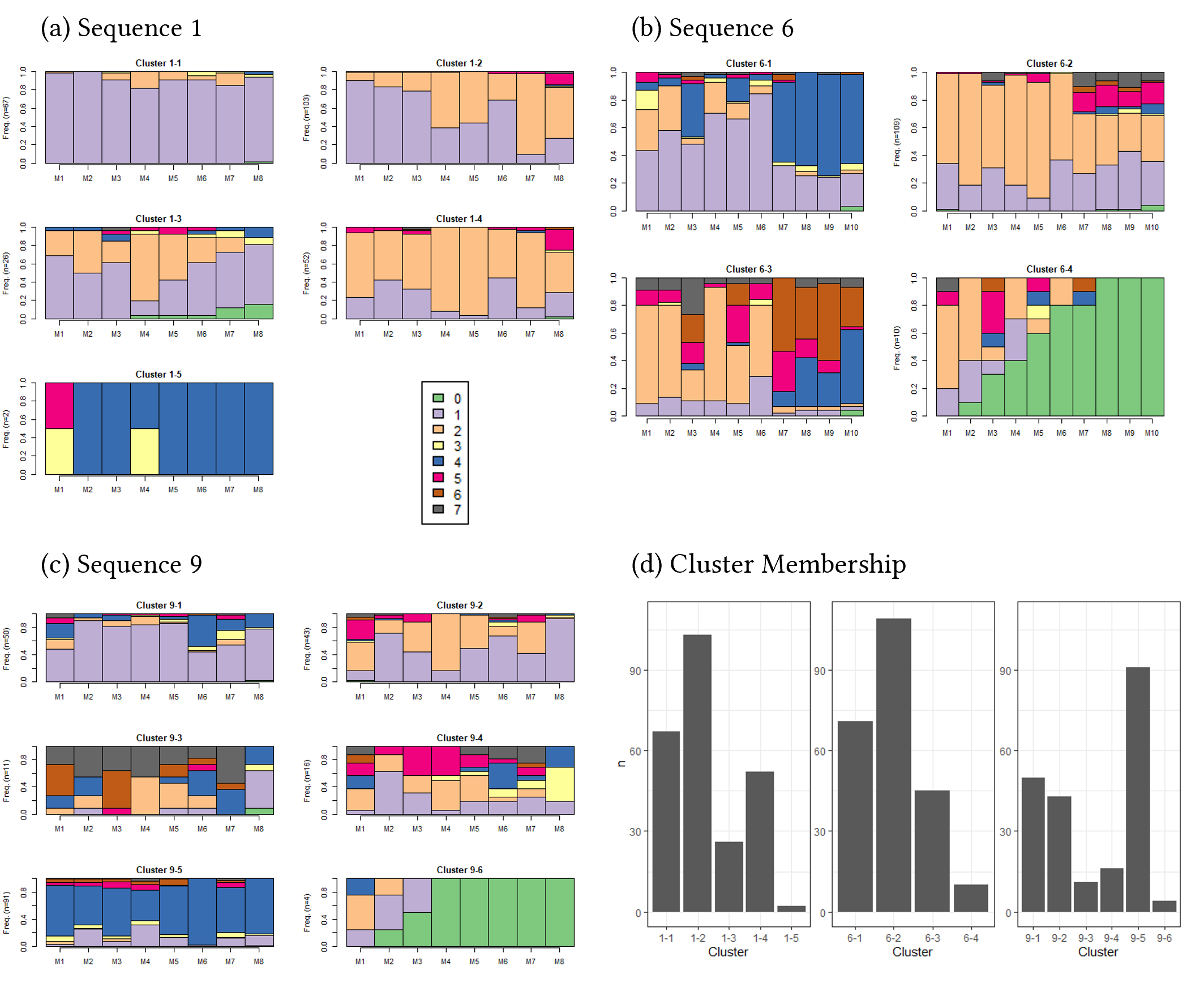}
    \caption{Visualization of the s-clusters of each sequence and overall cluster membership by sequence.}
    \label{fig: sClusters}
\end{figure}

A total of 5, 4, and 6 s-clusters are identified for sequences 1, 6, and 9 respectively according to the average silhouette scores. In figure \ref{fig: sClusters}, we visualize the characteristics of each identified s-cluster by plotting the distribution of m-clusters for each OLM in the sequence using stacked bar charts. Each bar chart corresponds to one identified s-cluster and each column corresponds to one OLM in the given sequence. The height of the bars represents the fraction of event traces that belong to one of the seven m-clusters. Here, m-cluster 0 is used to indicate that the student did not interact with the given module. Based on the frequency of observing different m-clusters within each sequence, the s-clusters can be roughly categorized into five types.

\textbf{\textit{Type I: Initial Pass}}. Traces classified into s-clusters 1-1, 6-1, and 9-1, share the common feature of having predominantly m-cluster 1 (A) on all or most of the modules. S-clusters 6-1 and 9-1 also had a smaller fraction of m-cluster 4 (S) which in the case of s-cluster 6-1 became dominant in the last 4 modules.

\textbf{\textit{Type II: Pass or Study}}. Traces classified into s-clusters 1-2, 1-3, 1-4, 6-2, and 9-2 share the common feature of having a combination of m-clusters 1 (A) and 2 (AYAn). S-clusters 1-2 and 1-3 are different in that 1-2 had higher fractions of m-cluster 1 (A) on the first couple of modules whereas the distribution in 1-3 is roughly uniform and a bit higher in the later modules. S-cluster 1-3 is also smaller than 1-2 and 1-4. S-cluster 1-4 consists of predominantly m-cluster 2 (AYAn) across all modules, which is also the case for s-cluster 6-2. However, on the last 4 modules, of s-cluster 6-2, a significantly higher fraction of m-clusters 5, 6, and 7 are observed.

\textbf{\textit{Type III: Varied Strategies}}. For traces in s-clusters 6-3, 9-3, and 9-4, there isn't a  single dominant m-cluster, but m-clusters 3 (Sn), 5 (AYSn), and 7 (SYAn) are observed at a higher frequency when compared to other s-clusters. m-clusters 5, 6, and 7 all involve both a study event followed by one or more short attempts. A notable difference between those three clusters is that s-cluster 6-3 has a higher fraction of m-cluster 6 (SYSn) and 4 (S) on the last four modules, suggesting ineffective study or giving up without studying. S-clusters 9-3 has a higher fraction of m-cluster 7 (SYAn), which is study and normal attempts after an initial guessing attempt on the mandatory first attempt.

\textbf{\textit{Type IV: Short Pass}}. Traces in s-clusters 1-5 and 9-5 consist of predominantly, m-cluster 4 (S),  short passing on the first attempt, on all modules in the sequence, which strongly suggests some form of answer copying or quick guessing. While s-cluster 1-5 consists of only 2 students, s-cluster 9-5 consists of 91 students and is the single largest cluster in sequence 9.

\textbf{\textit{Type V: Incomplete}}. The small number of traces in s-clusters 6-4 and 9-6 do not have records on the last couple of modules in the sequences. Note that there are also several unfinished traces in cluster 1-3, but the missing data is predominantly concentrated on the last couple of modules. 

\subsection{S-cluster membership and Assessment Performance}
\begin{figure}
    \centering
    \includegraphics[width = .75\textwidth]{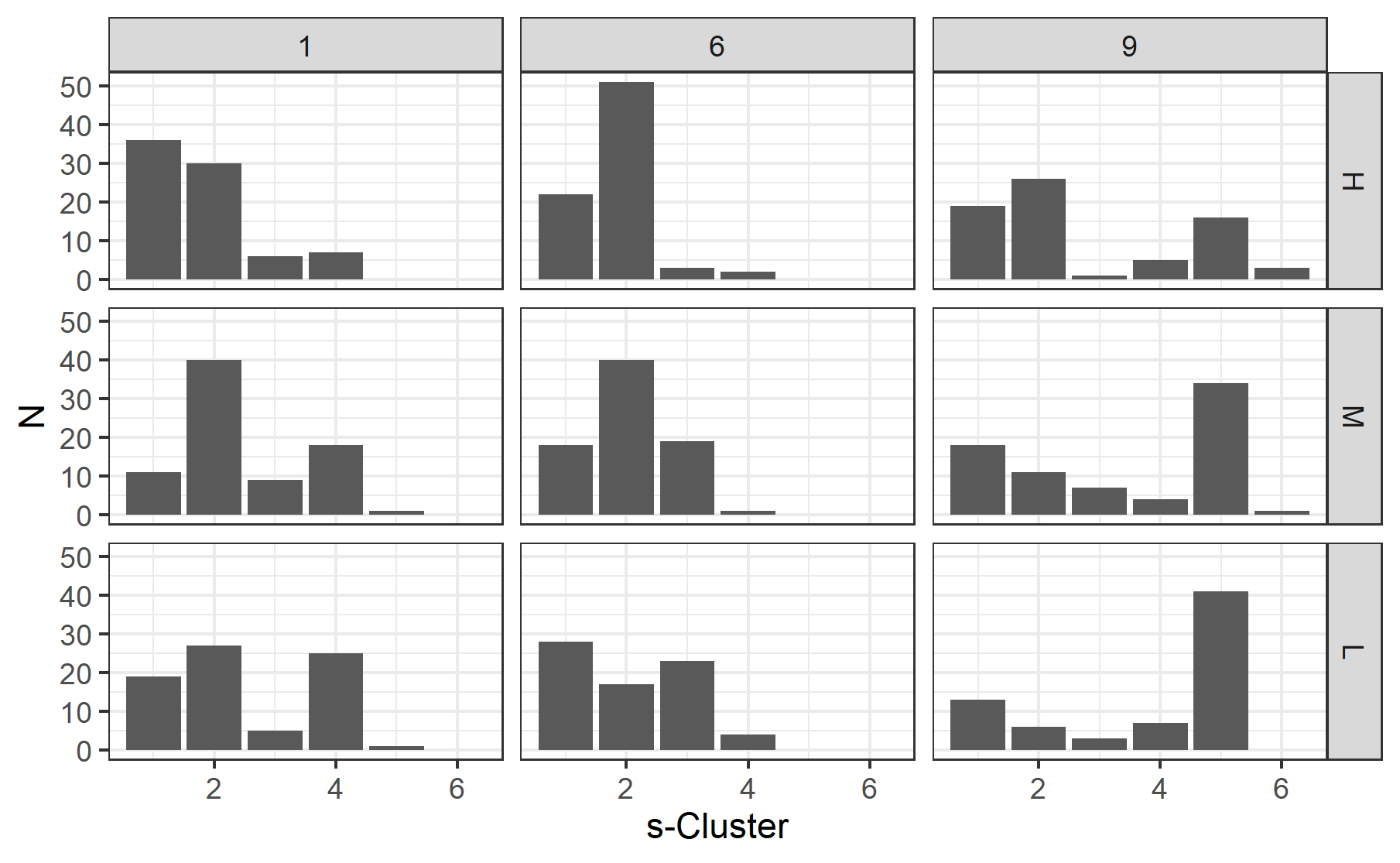}
    \caption{S-cluster membership for each sequence and quiz score tertile cohort.}
    \label{fig: Quiz Cluster Membership}
\end{figure}

Fisher tests conducted for each sequence detected significant differences in s-cluster membership among high, medium and low quiz score cohorts ($p < 0.0001$ for each sequence). For each sequence, the distribution of s-cluster memberships among each quiz score cohort is plotted in figure \ref{fig: Quiz Cluster Membership} whereas the result of post hoc analysis is listed in table \ref{tab: Assessment}

\begin{table}[!ht]
    \centering
    \begin{tabular}{c c c c}
        \toprule
        \toprule
        s-cluster & H:M & H:L & M:L \\
        \midrule
        1-1 & 0.0003*** & 0.037* & ns \\
        1-4 & ns & 0.0022** & ns \\
        6-2 & ns & 0.000003*** & 0.002** \\
        6-3 & 0.0013* & 0.00003*** & ns \\
        9-2 & 0.014* & 0.0008*** & ns \\
        9-5 & 0.024* & 0.00053*** & ns\\
        \bottomrule
        \bottomrule
    \end{tabular}
    \caption{Adjusted p-values for significant differences in s-cluster membership between average quiz score tertiles.}
    \label{tab: Assessment}
\end{table}

\section{Discussion}\label{sec: Discussion}
The multi-level hierarchical clustering analysis scheme produced relatively well-organized and highly interpretable clusters at both the module level and the sequence level. As one of the centerpieces of the analysis, the purely theory-based distance metric at the module level (Level II) outperformed both data-based and hybrid distance metrics, by producing both better-structured clusters (as judged by average silhouette) and more interpretable outcomes under the framework of SRL (RQ1). A possible implication is that in a restrictive system such as the OLMs, instructional conditions imposed by the system design play a major role in students' decision-making process, but may not be well reflected in the naturally occurring event frequencies. Therefore a strong input of expert knowledge is required to produce well-organized and interpretable clusters. 

Results from module-level clustering answered RQ2 by revealing three major m-clusters accounting for about 86\% of all traces, together with four smaller m-clusters. The largest m-cluster 1 (A), passing on the initial attempt before study, could indicate that a student entered the OLM with high prior content knowledge, possibly from learning the content from video lectures or in previous courses. However, it is also possible that in some cases students might have engaged in undesirable behaviors like conducting internet searches for problem solutions, which is indistinguishable from a normal attempt based on duration data alone. The second-largest m-cluster is 2 (AYAn), passing after study. Both m-clusters 1 and 2 represent the kinds of effective self-regulatory student behaviors that the OLMs were designed to elicit. The majority (74.3\%) of study events in this cluster took place immediately after the first attempt, showing that students either actively reflected on their problem solving-performance, or planned their learning strategy according to the instructor's recommendation. The third major m-cluster is 4 (S), pass on one brief attempt cluster, which suggests possible answer copying and counter-productive self-regulation during learning. These three m-clusters represent learning strategies frequently observed in all three sequences, while the four minor m-clusters reflect strategies mostly adopted by some students on specific modules and sequences. 

The outcomes of the sequence level clustering provided the most insight into students' SRL strategies across multiple modules and how those strategies changed over time (RQ3). As shown in Figure \ref{fig: sClusters}, learning strategies employed by students changed significantly across the three sequences. For sequences 1 and 6, the most populous s-clusters were 1-2 and 6-2, both of which contain a mixture of m-clusters 1 (A) and 2 (AYAn), indicating effective strategy use as intended by the OLM design.  However, by the end of the course, the dominant s-cluster for sequence 9 is 9-4, which almost exclusively consists of m-cluster 4 (S), indicating consistent and deliberate answer copying. Strikingly, the same "copy or guess" strategy was only adopted by 2 students in sequence 1 in s-cluster 1-5.

Students also changed their learning strategies within a single OLM sequence. For example in sequence 6, a shift in strategy is observed between modules 6 and 7 for all s-clusters. While the dominant strategy for the first 5 OLMs consists of m-clusters 1 (A) and 2 (AYAn), there was an obvious increase in m-clusters 4 (S), 5(AYSn), and 6 (SYSn) on the last 4 modules. Both m-clusters 5 and 6 are characterized by finishing on multiple short attempts following a study event, which could indicate that those students had difficulty learning from the instructional materials and therefore resorted to guessing on subsequent attempts. This shift in strategy could be explained in part by a change in content difficulty, as OLMs 1-6 focused on introducing individual concepts, whereas M7-M10 focused on the synthesis of multiple concepts and mathematical modeling. 

Moreover, a smaller fraction of students displayed unique learning strategies towards the end of the semester when the content became even harder. In s-cluster 9-3 for example, students consistently adopted m-clusters 6 (SYSn) and 7 (SYAn). A similar trend was also observed for the last four modules in s-clusters 6-2 and 6-3. Both m-clusters are characterized by a short first attempt, followed by a study event, which could be interpreted as students strategically skipping the required first attempt, likely due to a lack of self-confidence in their ability to solve the problem. Right after the first guessing attempt, those students followed their plan to study the instructional materials, with the outcome being a mix of success (7) and failure (6).  
		 
The correlation between s-cluster membership and course assessment performance provides further evidence for the interpretation of s-clusters. The three s-clusters that are more frequently observe among the top or middle tertile, 1-1, 6-2, and 9-3, all consist of predominantly m-clusters 1 (A) and 2 (AYAn).  On the other hand, the two s-clusters, 6-3 and 9-5, involving a significant fraction of m-clusters 4 (S), 5 (AYSn), and 6 (SYSn), are significantly more frequently observed among the bottom tertile. The one exception being 1-4, which is predominantly m-cluster 2, yet contains significantly more low performers. Consistent with existing research on incoming knowledge \cite{Incoming1}, students who do not have prior knowledge of the content in the first week of the course generally performed lower on assessments.

The current study makes several contributions both in terms of analysis methodology for trace data and new insights into students' SRL behavior in an online learning environment. Regarding analysis methodology, we overcame the challenges imposed by a more limited and restrictive online system with three main innovations. First, key information contained within the duration of events is extracted using mixture modeling. Second, we used an SRL theoretical framework to inform the clustering algorithm of pedagogically important features in trace data and showed that it outperformed more conventional data-based sequence-mining techniques. Finally, we used sequence pattern analysis in the final stage of clustering to reveal information on students' change of learning strategies over time. 

Regarding the insights gained on students' SRL behavior, the most important takeaway from the current analysis is that the majority of students are continuously self-regulating and adjusting their strategies from one module or one sequence to the next. We did not find a group of students whose behavior consistently indicates a lack of self-regulation. Rather, even when a student employs counterproductive strategies such as answer copying or guessing, it is more likely the result of a deliberate choice in response to challenges such as increased content difficulty or fatigue towards the end of the semester. For instructors, this observation reminds us that rather than blaming students for lack of academic integrity or self-discipline, it is essential to re-examine existing instructional design. It could be beneficial to pick up the pace early on when most students are actively engaged, and space out the more challenging content towards the end to allow for more opportunity to overcome challenges. Moreover, it seems that the prime time to implement interventions directed at improving students' SRL skills is the middle or latter half of the semester, rather than at the beginning. 

\subsection{Caveats and Future Directions}

From a technical standpoint, the current m-clusters reflected planning and decision-making more than learning outcomes, as the same m-cluster includes traces that pass on either 2 or 5 attempts. This could be caused by the choice of feature weights for the Gower coefficients, but also by the restriction of having a maximum of 10 clusters. In addition to conducting a more systematic search of the clustering parameter space, future studies could explore whether there can actually be more than 10 meaningful clusters, or that the trace data can be clustered in more than one way.
In addition, future studies could employ process mining algorithms to discover whether there exist less frequent processes in each m-cluster. An exploratory analysis employing heuristic miner algorithm \cite{heuristicsmineR} on 80\% of traces created causal nets that are highly similar to the current process maps, but with some non-trivial differences in less frequent paths and states.

From a pedagogical standpoint, future studies will need to investigate students' different strategies interacting with the instructional materials and practice problems, which in the current study is simplified into a single binary variable. Moreover, important insight into students' learning behavior can be gained by tracing how students change strategies from one sequence to the next over the semester, especially when combined with self-report data such as surveys on students' SRL strategies and goals and orientations.

Finally, a highly valuable future direction would be to combine or integrate the current multi-level clustering analysis scheme with other schemes such as Trace-SRL, to study students' SRL behavior in an open online learning environment consisting of multiple restrictive sub-systems.  

\begin{acks}
This work is supported by NSF Award No. DUE-1845436. We would like to thank UCF Center for Distributed Learning for creating the UCF Open project and the Obojobo platform, in particular Dr. Francisca Yonekura, Ian Turgeon, Allison Banzon, and Zachary Berry. 
\end{acks}
\bibliographystyle{ACM-Reference-Format}
\bibliography{bibliography}


\begin{thebibliography}{28}


\ifx \showCODEN    \undefined \def \showCODEN     #1{\unskip}     \fi
\ifx \showDOI      \undefined \def \showDOI       #1{#1}\fi
\ifx \showISBNx    \undefined \def \showISBNx     #1{\unskip}     \fi
\ifx \showISBNxiii \undefined \def \showISBNxiii  #1{\unskip}     \fi
\ifx \showISSN     \undefined \def \showISSN      #1{\unskip}     \fi
\ifx \showLCCN     \undefined \def \showLCCN      #1{\unskip}     \fi
\ifx \shownote     \undefined \def \shownote      #1{#1}          \fi
\ifx \showarticletitle \undefined \def \showarticletitle #1{#1}   \fi
\ifx \showURL      \undefined \def \showURL       {\relax}        \fi
\providecommand\bibfield[2]{#2}
\providecommand\bibinfo[2]{#2}
\providecommand\natexlab[1]{#1}
\providecommand\showeprint[2][]{arXiv:#2}

\bibitem[\protect\citeauthoryear{Benjamini and Yekutieli}{Benjamini and
  Yekutieli}{2001}]%
        {fdr}
\bibfield{author}{\bibinfo{person}{Yoav Benjamini} {and}
  \bibinfo{person}{Daniel Yekutieli}.} \bibinfo{year}{2001}\natexlab{}.
\newblock \showarticletitle{{The control of the false discovery rate in
  multiple testing under dependency}}.
\newblock \bibinfo{journal}{\emph{The Annals of Statistics}}
  \bibinfo{volume}{29}, \bibinfo{number}{4} (\bibinfo{year}{2001}),
  \bibinfo{pages}{1165 -- 1188}.
\newblock
\urldef\tempurl%
\url{https://doi.org/10.1214/aos/1013699998}
\showDOI{\tempurl}


\bibitem[\protect\citeauthoryear{Bloom}{Bloom}{1968}]%
        {Bloom1968}
\bibfield{author}{\bibinfo{person}{B Bloom}.} \bibinfo{year}{1968}\natexlab{}.
\newblock \showarticletitle{{Learning for Mastery. Instruction and Curriculum.
  Regional Education Laboratory for the Carolinas and Virginia, Topical Papers
  and Reprints, Number 1.}}
\newblock \bibinfo{journal}{\emph{Evaluation comment}}  \bibinfo{volume}{1}
  (\bibinfo{year}{1968}), \bibinfo{pages}{12}.
\newblock


\bibitem[\protect\citeauthoryear{{Center for Distributed Learning}}{{Center for
  Distributed Learning}}{[n.d.]}]%
        {CenterforDistributedLearning}
\bibfield{author}{\bibinfo{person}{{Center for Distributed Learning}}.}
  \bibinfo{year}{[n.d.]}\natexlab{}.
\newblock \bibinfo{title}{{Obojobo}}.
\newblock
\newblock


\bibitem[\protect\citeauthoryear{Chen, Xu, Garrido, and Guthrie}{Chen
  et~al\mbox{.}}{2020}]%
        {Chen2020}
\bibfield{author}{\bibinfo{person}{Zhongzhou Chen}, \bibinfo{person}{Mengyu
  Xu}, \bibinfo{person}{Geoffrey Garrido}, {and} \bibinfo{person}{Matthew~W.
  Guthrie}.} \bibinfo{year}{2020}\natexlab{}.
\newblock \showarticletitle{{Relationship between students' online learning
  behavior and course performance: What contextual information matters?}}
\newblock \bibinfo{journal}{\emph{Physical Review Physics Education Research}}
  \bibinfo{volume}{16}, \bibinfo{number}{1} (\bibinfo{date}{jun}
  \bibinfo{year}{2020}), \bibinfo{pages}{010138}.
\newblock
\showISSN{2469-9896}
\urldef\tempurl%
\url{https://doi.org/10.1103/PhysRevPhysEducRes.16.010138}
\showDOI{\tempurl}


\bibitem[\protect\citeauthoryear{Ericsson, Krampe, and
  Tesch-R{\"{o}}mer}{Ericsson et~al\mbox{.}}{1993}]%
        {Ericsson1993}
\bibfield{author}{\bibinfo{person}{K~Anders Ericsson}, \bibinfo{person}{Ralf~T
  Krampe}, {and} \bibinfo{person}{Clemens Tesch-R{\"{o}}mer}.}
  \bibinfo{year}{1993}\natexlab{}.
\newblock \showarticletitle{{The role of deliberate practice in the acquisition
  of expert performance.}}
\newblock \bibinfo{journal}{\emph{Psychological Review}} \bibinfo{volume}{100},
  \bibinfo{number}{3} (\bibinfo{year}{1993}), \bibinfo{pages}{363--406}.
\newblock
\showISSN{1939-1471(Electronic),0033-295X(Print)}
\urldef\tempurl%
\url{https://doi.org/10.1037/0033-295X.100.3.363}
\showDOI{\tempurl}


\bibitem[\protect\citeauthoryear{Fan, Saint, Singh, Jovanovic, and
  Ga{\v{s}}evi{\'{c}}}{Fan et~al\mbox{.}}{2021}]%
        {Fan2021}
\bibfield{author}{\bibinfo{person}{Yizhou Fan}, \bibinfo{person}{John Saint},
  \bibinfo{person}{Shaveen Singh}, \bibinfo{person}{Jelena Jovanovic}, {and}
  \bibinfo{person}{Dragan Ga{\v{s}}evi{\'{c}}}.}
  \bibinfo{year}{2021}\natexlab{}.
\newblock \showarticletitle{{A learning analytic approach to unveiling
  self-regulatory processes in learning tactics}}.
\newblock \bibinfo{journal}{\emph{ACM International Conference Proceeding
  Series}} (\bibinfo{year}{2021}), \bibinfo{pages}{184--195}.
\newblock
\showISBNx{9781450389358}
\urldef\tempurl%
\url{https://doi.org/10.1145/3448139.3448211}
\showDOI{\tempurl}


\bibitem[\protect\citeauthoryear{Felker and Chen}{Felker and Chen}{2020}]%
        {Felker2020}
\bibfield{author}{\bibinfo{person}{Zachary Felker} {and}
  \bibinfo{person}{Zhongzhou Chen}.} \bibinfo{year}{2020}\natexlab{}.
\newblock \showarticletitle{{The impact of extra credit incentives on students'
  work habits when completing online homework assignments}}. In
  \bibinfo{booktitle}{\emph{2020 Physics Education Research Conference
  Proceedings}}. \bibinfo{publisher}{American Association of Physics Teachers
  (AAPT)}, \bibinfo{address}{Virtual Conference}, \bibinfo{pages}{143--148}.
\newblock
\urldef\tempurl%
\url{https://doi.org/10.1119/perc.2020.pr.felker}
\showDOI{\tempurl}


\bibitem[\protect\citeauthoryear{Gabadinho, Ritschard, Müller, and
  Studer}{Gabadinho et~al\mbox{.}}{2011}]%
        {TraMineR}
\bibfield{author}{\bibinfo{person}{Alexis Gabadinho}, \bibinfo{person}{Gilbert
  Ritschard}, \bibinfo{person}{Nicolas~S. Müller}, {and}
  \bibinfo{person}{Matthias Studer}.} \bibinfo{year}{2011}\natexlab{}.
\newblock \showarticletitle{Analyzing and Visualizing State Sequences in {R}
  with {TraMineR}}.
\newblock \bibinfo{journal}{\emph{Journal of Statistical Software}}
  \bibinfo{volume}{40}, \bibinfo{number}{4} (\bibinfo{year}{2011}),
  \bibinfo{pages}{1--37}.
\newblock
\urldef\tempurl%
\url{https://doi.org/10.18637/jss.v040.i04}
\showDOI{\tempurl}


\bibitem[\protect\citeauthoryear{Garrido, Guthrie, and Chen}{Garrido
  et~al\mbox{.}}{2020}]%
        {Garrido2020}
\bibfield{author}{\bibinfo{person}{Geoffrey Garrido},
  \bibinfo{person}{Matthew~W. Guthrie}, {and} \bibinfo{person}{Zhongzhou
  Chen}.} \bibinfo{year}{2020}\natexlab{}.
\newblock \showarticletitle{{How are students' online learning behavior related
  to their course outcomes in an introductory physics course?}}. In
  \bibinfo{booktitle}{\emph{2019 Physics Education Research Conference
  Proceedings}}, \bibfield{editor}{\bibinfo{person}{Ying Cao},
  \bibinfo{person}{Steven Wolf}, {and} \bibinfo{person}{Michael~B. Bennett}}
  (Eds.). \bibinfo{publisher}{American Association of Physics Teachers},
  \bibinfo{address}{Provo, UT}.
\newblock
\urldef\tempurl%
\url{https://doi.org/10.1119/perc.2019.pr.Garrido}
\showDOI{\tempurl}


\bibitem[\protect\citeauthoryear{Guthrie, Zhang, and Chen}{Guthrie
  et~al\mbox{.}}{2020}]%
        {Guthrie2020}
\bibfield{author}{\bibinfo{person}{Matthew~W. Guthrie}, \bibinfo{person}{Tom
  Zhang}, {and} \bibinfo{person}{Zhongzhou Chen}.}
  \bibinfo{year}{2020}\natexlab{}.
\newblock \showarticletitle{{A tale of two guessing strategies: interpreting
  the time students spend solving problems through online log data}}. In
  \bibinfo{booktitle}{\emph{Physics Education Research Conference
  Proceedings}}. \bibinfo{publisher}{American Association of Physics Teachers
  (AAPT)}, \bibinfo{address}{Virtual Conference}, \bibinfo{pages}{185--190}.
\newblock
\urldef\tempurl%
\url{https://doi.org/10.1119/perc.2020.pr.guthrie}
\showDOI{\tempurl}


\bibitem[\protect\citeauthoryear{Gutmann, Gladding, Lundsgaard, and
  Stelzer}{Gutmann et~al\mbox{.}}{[n.d.]}]%
        {Gutmann}
\bibfield{author}{\bibinfo{person}{Brianne Gutmann}, \bibinfo{person}{Gary~E.
  Gladding}, \bibinfo{person}{Morten Lundsgaard}, {and}
  \bibinfo{person}{Timothy Stelzer}.} \bibinfo{year}{[n.d.]}\natexlab{}.
\newblock \showarticletitle{{Mastery-style homework exercises in introductory
  physics courses: Implementation matters}}.
\newblock \bibinfo{journal}{\emph{Physical Review Physics Education Research}}
  (\bibinfo{year}{[n.\,d.]}).
\newblock


\bibitem[\protect\citeauthoryear{Janssenswillen}{Janssenswillen}{2020a}]%
        {bupaR}
\bibfield{author}{\bibinfo{person}{Gert Janssenswillen}.}
  \bibinfo{year}{2020}\natexlab{a}.
\newblock \bibinfo{booktitle}{\emph{bupaR: Business Process Analysis in R}}.
\newblock
\urldef\tempurl%
\url{https://CRAN.R-project.org/package=bupaR}
\showURL{%
\tempurl}
\newblock
\shownote{R package version 0.4.4.}


\bibitem[\protect\citeauthoryear{Janssenswillen}{Janssenswillen}{2020b}]%
        {processmapR}
\bibfield{author}{\bibinfo{person}{Gert Janssenswillen}.}
  \bibinfo{year}{2020}\natexlab{b}.
\newblock \bibinfo{booktitle}{\emph{processmapR: Construct Process Maps Using
  Event Data}}.
\newblock
\urldef\tempurl%
\url{https://CRAN.R-project.org/package=processmapR}
\showURL{%
\tempurl}
\newblock
\shownote{R package version 0.3.4.}


\bibitem[\protect\citeauthoryear{Jovanovi{\'{c}}, Ga{\v{s}}evi{\'{c}}, Dawson,
  Pardo, and Mirriahi}{Jovanovi{\'{c}} et~al\mbox{.}}{2017}]%
        {Jovanovic2017}
\bibfield{author}{\bibinfo{person}{Jelena Jovanovi{\'{c}}},
  \bibinfo{person}{Dragan Ga{\v{s}}evi{\'{c}}}, \bibinfo{person}{Shane Dawson},
  \bibinfo{person}{Abelardo Pardo}, {and} \bibinfo{person}{Negin Mirriahi}.}
  \bibinfo{year}{2017}\natexlab{}.
\newblock \showarticletitle{{Learning analytics to unveil learning strategies
  in a flipped classroom}}.
\newblock \bibinfo{journal}{\emph{Internet and Higher Education}}
  \bibinfo{volume}{33} (\bibinfo{year}{2017}), \bibinfo{pages}{74--85}.
\newblock
\showISSN{10967516}
\urldef\tempurl%
\url{https://doi.org/10.1016/j.iheduc.2017.02.001}
\showDOI{\tempurl}


\bibitem[\protect\citeauthoryear{Maechler, Rousseeuw, Struyf, Hubert, and
  Hornik}{Maechler et~al\mbox{.}}{2021}]%
        {cluster}
\bibfield{author}{\bibinfo{person}{Martin Maechler}, \bibinfo{person}{Peter
  Rousseeuw}, \bibinfo{person}{Anja Struyf}, \bibinfo{person}{Mia Hubert},
  {and} \bibinfo{person}{Kurt Hornik}.} \bibinfo{year}{2021}\natexlab{}.
\newblock \bibinfo{booktitle}{\emph{cluster: Cluster Analysis Basics and
  Extensions}}.
\newblock
\urldef\tempurl%
\url{https://CRAN.R-project.org/package=cluster}
\showURL{%
\tempurl}
\newblock
\shownote{R package version 2.1.2 --- For new features, see the 'Changelog'
  file (in the package source).}


\bibitem[\protect\citeauthoryear{Maldonado-Mahauad,
  P{\'{e}}rez-Sanagust{\'{i}}n, Kizilcec, Morales, and
  Munoz-Gama}{Maldonado-Mahauad et~al\mbox{.}}{2018}]%
        {Maldonado-Mahauad2018}
\bibfield{author}{\bibinfo{person}{Jorge Maldonado-Mahauad},
  \bibinfo{person}{Mar P{\'{e}}rez-Sanagust{\'{i}}n},
  \bibinfo{person}{Ren{\'{e}}~F. Kizilcec}, \bibinfo{person}{Nicol{\'{a}}s
  Morales}, {and} \bibinfo{person}{Jorge Munoz-Gama}.}
  \bibinfo{year}{2018}\natexlab{}.
\newblock \showarticletitle{{Mining theory-based patterns from Big data:
  Identifying self-regulated learning strategies in Massive Open Online
  Courses}}.
\newblock \bibinfo{journal}{\emph{Computers in Human Behavior}}
  \bibinfo{volume}{80} (\bibinfo{year}{2018}), \bibinfo{pages}{179--196}.
\newblock
\showISSN{07475632}
\urldef\tempurl%
\url{https://doi.org/10.1016/j.chb.2017.11.011}
\showDOI{\tempurl}


\bibitem[\protect\citeauthoryear{Mannhardt}{Mannhardt}{2020}]%
        {heuristicsmineR}
\bibfield{author}{\bibinfo{person}{Felix Mannhardt}.}
  \bibinfo{year}{2020}\natexlab{}.
\newblock \bibinfo{booktitle}{\emph{heuristicsmineR: Discovery of Process
  Models with the Heuristics Miner}}.
\newblock
\urldef\tempurl%
\url{https://CRAN.R-project.org/package=heuristicsmineR}
\showURL{%
\tempurl}
\newblock
\shownote{R package version 0.2.4.}


\bibitem[\protect\citeauthoryear{Prates, Cabral, and Lachos}{Prates
  et~al\mbox{.}}{2013}]%
        {mixsmsn}
\bibfield{author}{\bibinfo{person}{Marcos~Oliveira Prates},
  \bibinfo{person}{Celso R{\^o}mulo~Barbosa Cabral}, {and}
  \bibinfo{person}{V{\'i}ctor~Hugo Lachos}.} \bibinfo{year}{2013}\natexlab{}.
\newblock \showarticletitle{{mixsmsn}: Fitting Finite Mixture of Scale Mixture
  of Skew-Normal Distributions}.
\newblock \bibinfo{journal}{\emph{Journal of Statistical Software}}
  \bibinfo{volume}{54}, \bibinfo{number}{12} (\bibinfo{year}{2013}),
  \bibinfo{pages}{1--20}.
\newblock
\urldef\tempurl%
\url{https://www.jstatsoft.org/v54/i12/}
\showURL{%
\tempurl}


\bibitem[\protect\citeauthoryear{Rousseeuw}{Rousseeuw}{1987}]%
        {Silhouette}
\bibfield{author}{\bibinfo{person}{Peter~J. Rousseeuw}.}
  \bibinfo{year}{1987}\natexlab{}.
\newblock \showarticletitle{Silhouettes: A graphical aid to the interpretation
  and validation of cluster analysis}.
\newblock \bibinfo{journal}{\emph{J. Comput. Appl. Math.}}
  \bibinfo{volume}{20} (\bibinfo{year}{1987}), \bibinfo{pages}{53--65}.
\newblock
\showISSN{0377-0427}
\urldef\tempurl%
\url{https://doi.org/10.1016/0377-0427(87)90125-7}
\showDOI{\tempurl}


\bibitem[\protect\citeauthoryear{{Saint}, {Whitelock-Wainwright}, {Gašević},
  and {Pardo}}{{Saint} et~al\mbox{.}}{2020}]%
        {Saint2020}
\bibfield{author}{\bibinfo{person}{J. {Saint}}, \bibinfo{person}{A.
  {Whitelock-Wainwright}}, \bibinfo{person}{D. {Gašević}}, {and}
  \bibinfo{person}{A. {Pardo}}.} \bibinfo{year}{2020}\natexlab{}.
\newblock \showarticletitle{Trace-SRL: A Framework for Analysis of Microlevel
  Processes of Self-Regulated Learning From Trace Data}.
\newblock \bibinfo{journal}{\emph{IEEE Transactions on Learning Technologies}}
  \bibinfo{volume}{13}, \bibinfo{number}{4} (\bibinfo{date}{Oct}
  \bibinfo{year}{2020}), \bibinfo{pages}{861--877}.
\newblock
\showISSN{1939-1382}
\urldef\tempurl%
\url{https://doi.org/10.1109/TLT.2020.3027496}
\showDOI{\tempurl}


\bibitem[\protect\citeauthoryear{Salehi, Burkholder, Lepage, Pollock, and
  Wieman}{Salehi et~al\mbox{.}}{2019}]%
        {Incoming1}
\bibfield{author}{\bibinfo{person}{Shima Salehi}, \bibinfo{person}{Eric
  Burkholder}, \bibinfo{person}{G.~Peter Lepage}, \bibinfo{person}{Steven
  Pollock}, {and} \bibinfo{person}{Carl Wieman}.}
  \bibinfo{year}{2019}\natexlab{}.
\newblock \showarticletitle{Demographic gaps or preparation gaps?: The large
  impact of incoming preparation on performance of students in introductory
  physics}.
\newblock \bibinfo{journal}{\emph{Phys. Rev. Phys. Educ. Res.}}
  \bibinfo{volume}{15} (\bibinfo{date}{Jul} \bibinfo{year}{2019}),
  \bibinfo{pages}{020114}.
\newblock
Issue 2.
\urldef\tempurl%
\url{https://doi.org/10.1103/PhysRevPhysEducRes.15.020114}
\showDOI{\tempurl}


\bibitem[\protect\citeauthoryear{Schwartz and Bransford}{Schwartz and
  Bransford}{2005}]%
        {Schwartz2005}
\bibfield{author}{\bibinfo{person}{Daniel~L. Schwartz} {and}
  \bibinfo{person}{John~D. Bransford}.} \bibinfo{year}{2005}\natexlab{}.
\newblock \showarticletitle{{Efficiency and Innovation in Transfer}}.
\newblock In \bibinfo{booktitle}{\emph{Transfer of Learning from a Modern
  Multidisciplinary Perspective (Current Perspectives on Cognition, Learning
  and Instruction)}}, \bibfield{editor}{\bibinfo{person}{Jose~P. Mestre}}
  (Ed.). \bibinfo{publisher}{IAP - Informaiton Age Publishing Inc.},
  \bibinfo{address}{Charlotte}, \bibinfo{pages}{1--51}.
\newblock
\showISBNx{1593111649}


\bibitem[\protect\citeauthoryear{Whitcomb, Guthrie, Singh, and Chen}{Whitcomb
  et~al\mbox{.}}{2021}]%
        {Whitcomb2021}
\bibfield{author}{\bibinfo{person}{Kyle~M. Whitcomb},
  \bibinfo{person}{Matthew~W. Guthrie}, \bibinfo{person}{Chandralekha Singh},
  {and} \bibinfo{person}{Zhongzhou Chen}.} \bibinfo{year}{2021}\natexlab{}.
\newblock \showarticletitle{{Improving accuracy in measuring the impact of
  online instruction on students' ability to transfer physics problem-solving
  skills}}.
\newblock \bibinfo{journal}{\emph{Physical Review Physics Education Research}}
  \bibinfo{volume}{17}, \bibinfo{number}{1} (\bibinfo{date}{mar}
  \bibinfo{year}{2021}), \bibinfo{pages}{010112}.
\newblock
\showISSN{2469-9896}
\urldef\tempurl%
\url{https://doi.org/10.1103/PhysRevPhysEducRes.17.010112}
\showDOI{\tempurl}


\bibitem[\protect\citeauthoryear{Winne}{Winne}{2018}]%
        {Winne2018}
\bibfield{author}{\bibinfo{person}{Philip~H Winne}.}
  \bibinfo{year}{2018}\natexlab{}.
\newblock \showarticletitle{{Cognition and Metacognition within Self-Regulated
  Learning}}.
\newblock In \bibinfo{booktitle}{\emph{Handbook of self-regulation of learning
  and performance (2nd ed.)} (\bibinfo{edition}{2nd} ed.)},
  \bibfield{editor}{\bibinfo{person}{Patricia~A. Alexander},
  \bibinfo{person}{Dale~H. Schunk}, {and} \bibinfo{person}{Jeffrey~A. Greene}}
  (Eds.). Number 10449. \bibinfo{publisher}{Routledge}, \bibinfo{address}{New
  York}, Chapter Cognition, \bibinfo{pages}{36--48}.
\newblock
\showISBNx{9781315697048}
\urldef\tempurl%
\url{https://doi.org/10.4324/9781315697048.ch3}
\showDOI{\tempurl}


\bibitem[\protect\citeauthoryear{Zhang, Taub, and Chen}{Zhang
  et~al\mbox{.}}{2021}]%
        {Zhang2021}
\bibfield{author}{\bibinfo{person}{Tom Zhang}, \bibinfo{person}{Michelle Taub},
  {and} \bibinfo{person}{Zhongzhou Chen}.} \bibinfo{year}{2021}\natexlab{}.
\newblock \showarticletitle{Measuring the Impact of COVID-19 Induced Campus
  Closure on Student Self-Regulated Learning in Physics Online Learning
  Modules}. In \bibinfo{booktitle}{\emph{LAK21: 11th International Learning
  Analytics and Knowledge Conference}} (Irvine, CA, USA)
  \emph{(\bibinfo{series}{LAK21})}. \bibinfo{publisher}{Association for
  Computing Machinery}, \bibinfo{address}{New York, NY, USA},
  \bibinfo{pages}{110–120}.
\newblock
\showISBNx{9781450389358}
\urldef\tempurl%
\url{https://doi.org/10.1145/3448139.3448150}
\showDOI{\tempurl}


\bibitem[\protect\citeauthoryear{Zimmerman}{Zimmerman}{2000}]%
        {Zimmerman2000}
\bibfield{author}{\bibinfo{person}{Barry~J Zimmerman}.}
  \bibinfo{year}{2000}\natexlab{}.
\newblock \showarticletitle{{Attaining self-regulation: A social cognitive
  perspective.}}
\newblock In \bibinfo{booktitle}{\emph{Handbook of self-regulation.}}
  \bibinfo{publisher}{Academic Press}, \bibinfo{address}{San Diego},
  \bibinfo{pages}{13--39}.
\newblock
\showISBNx{0-12-109890-7 (Hardcover)}
\urldef\tempurl%
\url{https://doi.org/10.1016/B978-012109890-2/50031-7}
\showDOI{\tempurl}


\bibitem[\protect\citeauthoryear{Zimmerman}{Zimmerman}{2013}]%
        {Zimmerman2013}
\bibfield{author}{\bibinfo{person}{Barry~J. Zimmerman}.}
  \bibinfo{year}{2013}\natexlab{}.
\newblock \showarticletitle{{From Cognitive Modeling to Self-Regulation: A
  Social Cognitive Career Path}}.
\newblock \bibinfo{journal}{\emph{Educational Psychologist}}
  \bibinfo{volume}{48}, \bibinfo{number}{3} (\bibinfo{year}{2013}),
  \bibinfo{pages}{135--147}.
\newblock
\showISSN{00461520}
\urldef\tempurl%
\url{https://doi.org/10.1080/00461520.2013.794676}
\showDOI{\tempurl}


\bibitem[\protect\citeauthoryear{Zimmerman and Schunk}{Zimmerman and
  Schunk}{2011}]%
        {Zimmerman2011}
\bibfield{author}{\bibinfo{person}{Barry~J Zimmerman} {and}
  \bibinfo{person}{Dale~H Schunk}.} \bibinfo{year}{2011}\natexlab{}.
\newblock \showarticletitle{{Self-regulated learning and performance: An
  introduction and an overview.}}
\newblock In \bibinfo{booktitle}{\emph{Handbook of self-regulation of learning
  and performance.}} \bibinfo{publisher}{Routledge/Taylor {\&} Francis Group},
  \bibinfo{address}{New York}, \bibinfo{pages}{1--12}.
\newblock
\showISBNx{978-0-415-87112-9 (Paperback); 978-0-415-87111-2 (Hardcover);
  978-0-203-83901-0 (PDF)}


\end{thebibliography}

\end{document}